\documentclass[conference]{IEEEtran}
\IEEEoverridecommandlockouts
\usepackage{cite}
\usepackage{hyperref} 
\usepackage{amsmath,amssymb,amsfonts}
\usepackage{algorithmic}
\usepackage{graphicx}
\usepackage{textcomp}
\usepackage{xcolor}
\def\BibTeX{{\rm B\kern-.05em{\sc i\kern-.025em b}\kern-.08em
    T\kern-.1667em\lower.7ex\hbox{E}\kern-.125emX}}
\begin{document}

\title{DeepTelecom: A Digital-Twin Deep Learning Dataset for Channel and MIMO Applications\\
}

\author{
	\IEEEauthorblockN{Bohao Wang\IEEEauthorrefmark{1}\IEEEauthorrefmark{2}, Zehua Jiang\IEEEauthorrefmark{1}, Zhenyu Yang\IEEEauthorrefmark{6}, Chongwen Huang\IEEEauthorrefmark{1}\IEEEauthorrefmark{2}, Yongliang Shen\IEEEauthorrefmark{4}, Siming Jiang\IEEEauthorrefmark{9}, Chen Zhu\IEEEauthorrefmark{8},}
    
    \IEEEauthorblockN{Zhaohui Yang\IEEEauthorrefmark{1}, Richeng Jin\IEEEauthorrefmark{1}, Zhaoyang Zhang\IEEEauthorrefmark{1}, Sami Muhaidat\IEEEauthorrefmark{7}, and M\'{e}rouane~Debbah\IEEEauthorrefmark{5}\ \IEEEmembership{Fellow,~IEEE}}
	
	\IEEEauthorblockA{\IEEEauthorrefmark{1} College of Information Science and Electronic Engineering, Zhejiang University, 310027, Hangzhou, China}
	\IEEEauthorblockA{\IEEEauthorrefmark{2} State Key Laboratory of Integrated Service Networks, Xidian University, 710071, Xi'an, China}
	\IEEEauthorblockA{\IEEEauthorrefmark{6} School of Earth Sciences, Zhejiang University, 310027, Hangzhou, China}
	\IEEEauthorblockA{\IEEEauthorrefmark{4} College of Computer Science and Technology, Zhejiang University, 310027, Hangzhou, China}
	\IEEEauthorblockA{\IEEEauthorrefmark{9} Guangdong Tobacco Maoming Co., Ltd., 525000, Maoming, China}
	\IEEEauthorblockA{\IEEEauthorrefmark{8} Polytechnic Institute, Zhejiang University, 310015, Hangzhou, China}


    \IEEEauthorblockA{\IEEEauthorrefmark{7} Computer and Communication Engineering, Khalifa University, P.O. Box: 127788, Abu Dhabi, UAE}
    \IEEEauthorblockA{\IEEEauthorrefmark{5} KU 6G Research Center, Khalifa University of Science and Technology, P O Box 127788, Abu Dhabi, UAE}

    }
\maketitle

\begin{abstract}
Domain-specific datasets are the foundation for unleashing artificial intelligence (AI)-driven wireless innovation.  
Yet existing wireless AI corpora are slow to produce, offer limited modeling fidelity, and cover only narrow scenario types.
To address the challenges, we create DeepTelecom, a three-dimension (3D) digital-twin channel dataset. 
Specifically, a large language model (LLM)-assisted pipeline first builds the third level of details (LoD3) outdoor and indoor scenes with segmentable material-parameterizable surfaces. Then, DeepTelecom simulates full radio-wave propagation effects based on Sionna's ray-tracing engine. 
Leveraging GPU acceleration, DeepTelecom streams ray-path trajectories and real-time signal-strength heat maps, compiles them into high-frame-rate videos, and simultaneously outputs synchronized multi-view images, channel tensors, and multi-scale fading traces. 
By efficiently streaming large-scale, high-fidelity, and multimodal channel data, DeepTelecom not only furnishes a unified benchmark for wireless AI research but also supplies the domain-rich training substrate that enables foundation models to tightly fuse large model intelligence with future communication systems.
\end{abstract}

\begin{IEEEkeywords}
3D digital twin, large language model, ray-tracing, wireless AI, channel model.  
\end{IEEEkeywords}

\section{Introduction}
\begin{figure*}[h]
\centerline{\includegraphics[width=0.8\linewidth]{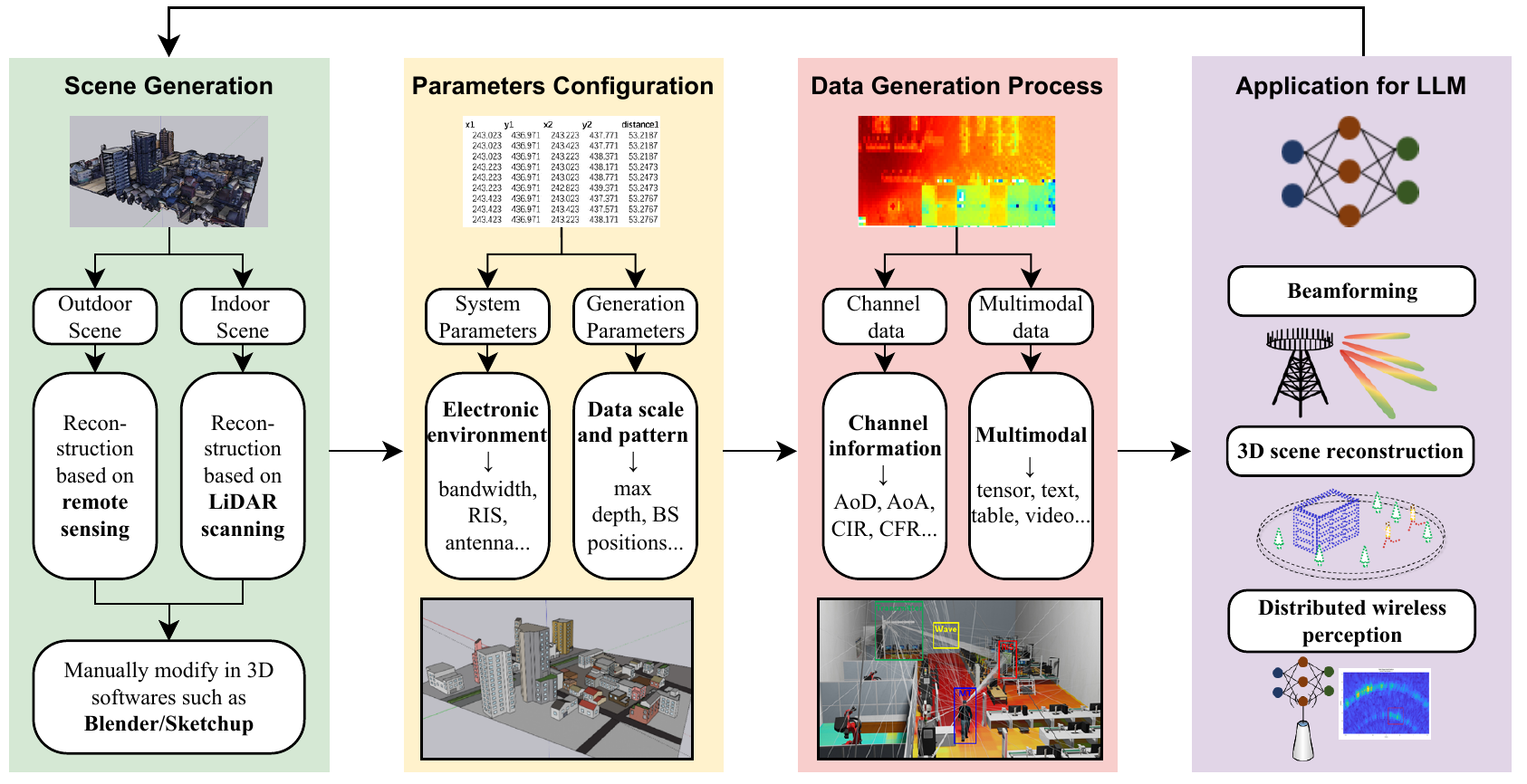}}
\caption{The overall DeepTelecom framework.}
\label{rayTracingFramework}
\vspace{-5mm}
\end{figure*}
Driven by the rapid rise of sixth-generation (6G) mobile communications, wireless systems are advancing toward ultra-high data rates, ultra-low latency, extreme reliability, and ubiquitous coverage.
Unlike fifth-generation (5G), 6G will serve as the core infrastructure of an intelligent society by embedding artificial intelligence (AI)-native, sub-THz, real-and-virtual co-designed signal processing at the physical layer.
AI, especially the fusion of large models (LMs) with multimodal learning, offers a new path to these goals by learning and reasoning over vast, heterogeneous spatiotemporal data, shifting 6G networks from connectivity toward artificial general intelligence \cite{shahid2025largescaleaitelecomcharting}.

Integrating AI and LMs into wireless communications depends fundamentally on building high-fidelity and large-scale training datasets \cite{zhu2025wirelesslargeaimodel}. 
Ray tracing remains the mainstream solution: in a three-dimension (3D) digital-twin scene, it virtually places base stations (BSs), mobile terminals (MTs), and reconfigurable intelligent surfaces (RISs), then deterministically traces electromagnetic paths to derive point-wise channel characteristics \cite{imai2017survey}.
Although industry and academic pioneers such as Huawei \cite{huangfu2022wair}, BUPT \cite{yu2024buptcmcc, shen2023dataai}, SEU’s PML-6GPCS simulator\cite{wang2022pervasive,wang2023complete}, and Peking University \cite{cheng2023m}, have piloted wireless-AI dataset programs, they still depend on CPU-bound ray tracers whose billion-ray runs are slow, modality-limited, and demand laborious calibration, resulting in digital-twin models confined to first level of details (LoD1) without per-material annotation.
By contrast, open or commercial toolchains such as DeepMIMO \cite{alkhateeb2019deepmimo}, ViWi \cite{Alrabeiah2023}, WiThRay \cite{choi2023withray}, Ranplan, and NVIDIA Sionna \cite{hoydis2022sionna} showcase GPU-accelerated, multi-band ray tracing and 3D rendering, yet none currently couples material-aware reconstruction with differentiable, physics-exact ray tracing in an integrated workflow.

To address these challenges, we generate DeepTelecom, a third level of details (LoD3) 3D digital wireless channel corpus that ingests openstreetmap and retains per-surface material semantics throughout the workflow.
Guided by a large language model (LLM) that converts text tags into frequency-dependent dielectric permittivity and magnetic permeability tensors, the engine parametrically configures transmitters, receivers, and RIS to systematically capture channel diversity across varying material properties and operational conditions.
It solves propagation via a hybrid geometrical-optics and geometrical-diffraction solver for specular, refractive, diffractive, and scattering effects based on Sionna differentiable GPU tracer, sustaining throughput beyond 10 million rays while preserving automatic-gradient capability. 
For every path the pipeline emits path-loss, delay, angle, phase, polarization, and doppler, concurrently rasterizes heat-maps and RGB overlays, and finally packages synchronized images, MP4 videos, HDF5 channel tensors, and CSV manifests under a unified scene index, thereby delivering the complete open multimodal dataset that fuses visual, tensor, and tabular views for end-to-end 6G AI research. 
The dataset and implementation are partially available online.\footnote{\url{https://project.veryengine.cn/publish/CGAQvTSeB/index.html}}

The rest of this paper is organized as follows. In Section II, we introduce the overall framework of the DeepTelecom framework. In Section III, we provide a detailed description of the implementation of its four core modules. Section IV demonstrates the generalizability of the generated scenes and the diversity of the resulting datasets. Finally, Section V concludes the paper.

\section{Framework Overview}
To create a high-fidelity, multi-modal, LoD3 GPU-accelerated ray-tracing dataset \textbf{DeepTelecom}, we divide the dataset generation workflow into four major components. The complete framework is shown in Fig. \ref{rayTracingFramework}, and each component is outlined as follows:
\begin{itemize}
    \item \textbf{LLM-Assisted High-Fidelity Multi-Source Scene Modeling:} We develop an LLM-assisted, high-precision scene modeling pipeline to construct LoD3 digital twins of both indoor and outdoor environments. This module integrates diverse data sources (e.g., satellite imagery, 3D datasets, and LiDAR scans) to produce realistic textures combined with accurate geometric structures, exporting to industry-standard 3D formats. The resulting indoor models capture intricate details of complex interiors, while outdoor models can cover urban areas with high fidelity. An LLM-based post-processing step then converts the reconstructed scene comprising all distinguishable objects  into a structured XML description. In this XML, each surface is explicitly defined and annotated with its electromagnetic material properties. This comprehensive scene representation closely mirrors real-world geometry and electromagnetic behavior, providing a solid foundation for subsequent ray-tracing simulations.
    
    \item \textbf{Parameter Definition and Configuration:} Following the environment modeling, we configure the simulation parameters, which are divided into two categories: \textit{system parameters} that define the hardware and devices, and \textit{generation parameters} that control the data generation process. The system configuration involves deploying BSs and MTs at specific 3D positions with defined orientations and transmit powers. To capture realistic channel dynamics, MTs' movement can be simulated and time-varing noise is introduced. Both BSs and MTs are equipped with configurable multiple-input multiple-output (MIMO) antenna arrays, whose size and element spacing are critical for accurately modeling spatial channel characteristics. Optionally, RIS can be integrated into the scene, configured for either single-beam focusing or multi-beamforming to intelligently control signal propagation. Finally, the system's bandwidth is defined, which determines the time and frequency resolution of the generated channel impulse response (CIR) and channel frequency response (CFR).
    
    \item \textbf{GPU-Accelerated Ray-Tracing Simulation:} We generate our channel data using a GPU-accelerated ray-tracing engine powered by NVIDIA's Sionna and OptiX libraries. This simulation is highly efficient, leveraging massive parallelization and optimization strategies like early ray termination (based on interaction depth and power thresholds). It accurately models key physical phenomena—including reflection, transmission, and diffraction—and records all significant path interactions to produce physically-grounded channel datasets.
    
    \item \textbf{Channel Data Extraction and Post-Processing:} From ray-traced paths, we derive the key channel metrics and compile the final dataset. Dedicated functions process the multipath data to compute standard channel representations including the CIR, CFR, and angular information like angle of arrival (AoA) and angle of departure (AoD) for each path. In this stage, we also perform final post-processing such as filtering out unreachable paths or invalid links to ensure data quality. The outcome is a comprehensive dataset of channel characteristics in each scenario, with high realism and diversity. 
\end{itemize}
In summary, DeepTelecom represents a significant advancement over existing channel datasets by uniquely combining large scale, scenario diversity, and high-fidelity digital-twin accuracy. It serves as a foundational asset for next-generation intelligent wireless communication research.

\section{Detailed Description}

In this section, we provide a detailed description of the DeepTelecom dataset generation process, expanding upon the four components outlined above. Each subsection explains how a particular part of the framework is implemented, including technical specifics and procedures.

\subsection{LLM-Assisted High-Fidelity Multi-Source Scene Modeling}

\subsubsection{Indoor Scene Reconstruction}

\begin{figure}[h]
\centerline{\includegraphics[width=1\linewidth]{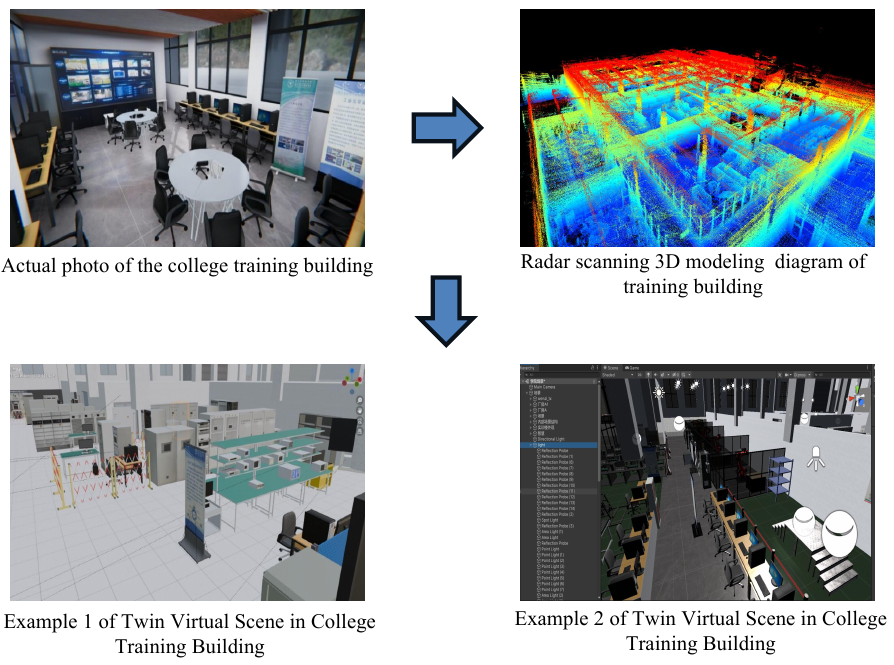}}
\caption{Indoor digital scene implementation diagram.}
\label{fig: indoor}
\vspace{-5mm}
\end{figure}

We utilize a cyber-physical digital twin system to reconstruct complex indoor scenes at LoD3 granularity. 
Firstly, a quadruped robot equipped with a 3D LiDAR scanner surveys the environment.
The LiDAR captures both geometric distances and reflection intensities, enabling inference of surface properties such as material type and color. 
The raw point cloud is then denoised, filtered, and converted into a unified 3D coordinate system. 
Here we apply an HDL-localization algorithm based on an unscented kalman filter (UKF) to achieve this unification by transforming each LiDAR point from the robot’s local frame to the global reference frame using the robot’s estimated pose \cite{zhu2024integrated}. 
Specifically, if at time $t$ the UKF provides a rotation $\mathbf{R}(t)$ and translation $\mathbf{t}(t)$ for the robot, a point measured in the robot’s local coordinates $\mathbf{p}_{\text{local}}$ is mapped to the global frame as: 
\begin{equation}
\mathbf{p}_{\text{global}}(t) = \mathbf{R}(t)\,\mathbf{p}_{\text{local}} + \mathbf{t}(t)\,,
\end{equation}
ensuring that all scans align in one common coordinate system. With the aggregated point cloud in this unified frame, the geometry and spatial layout of indoor structures like walls and furniture gradually emerge \cite{keller2013real}. The implementation diagram is shown in Fig.~\ref{fig: indoor}.

Once the comprehensive point cloud has been acquired, the system’s 3D modeling module converts it into a polygonal mesh and refines the model. We perform mesh extraction and simplification to reduce redundant geometry while preserving essential details. A combination of solid modeling, surface modeling, and wireframe modeling techniques is supported, catering to different reconstruction needs for various object types. 
We then apply texture baking and light baking, which pre-compute surface textures and lighting/shadow information and map them onto the mesh, yielding a photorealistic appearance. The final reconstructed indoor model, complete with high-resolution textures, can be exported in multiple formats (e.g., \texttt{.obj} or \texttt{.fbx}) for visualization and further use. This model serves as the digital twin of the real indoor space, capturing both its geometry and visual details.

\subsubsection{Outdoor Scene Reconstruction}

\begin{figure}[h]
\centerline{\includegraphics[width=1\linewidth]{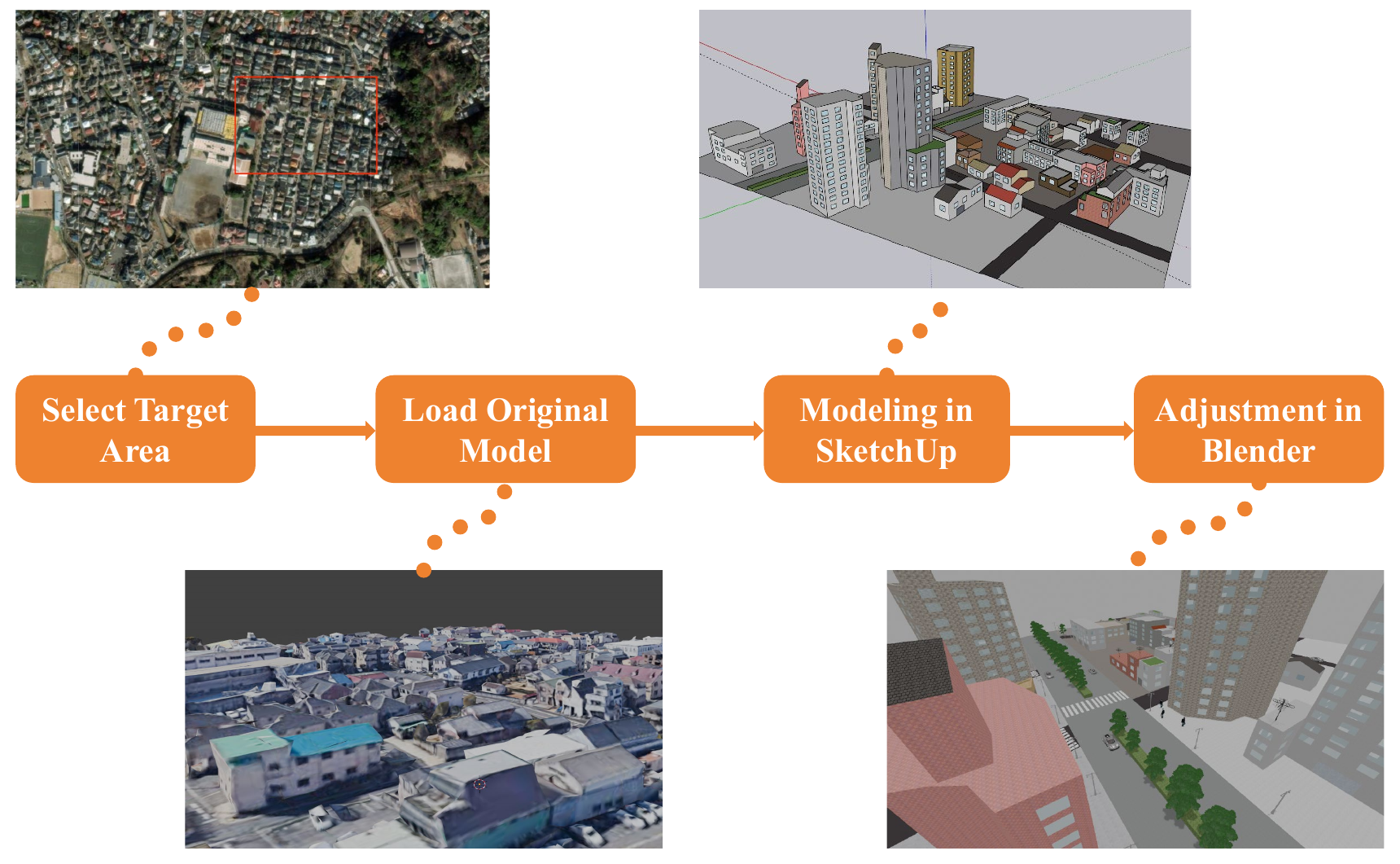}}
\caption{Outdoor digital scene implementation diagram.}
\label{fig:outdoor-scene}
\vspace{-2mm}
\end{figure}

For outdoor environments, we employ professional 3D modeling tools and open geospatial data to construct LoD3 urban scenes. 
The implementation example is depicted in Fig.~\ref{fig:outdoor-scene}.

In our pipeline, the Blender-Blosm plugin is used to import real-world models from Google 3D Tiles. A dense urban area in Japan is selected as the test site. Although these imported models include building shapes, textures, and terrain elevation, they are coarse, with uneven surfaces and merged components. 
For more efficient editting, we then export the models in \texttt{.gltf} format to SketchUp.
Through the application of modeling tools such as Push/Pull, we accurately rebuilt every key elements ensuring alignment with Google Tiles reference data, while systematically organizing objects by category and assigning realistic materials (e.g., glass, concrete) via SketchUp’s material editor.


\subsubsection{LLM-Assisted Scene Optimization}

After constructing the indoor and outdoor scene models, we export the full scene descriptions, including all objects and their materials, to a structured XML format for final refinement. 
At this stage, LLMs are used to optimize and validate the consistency of material annotations. Since LLMs are well-suited for structured text manipulation, they can effectively process the XML-based scene representation. To support this, we enforce a strict naming convention during modeling (e.g., “wall”, “window”, “door”, “roadSurface”). Using the XML file and a reference table of electromagnetic material properties, we prompt the LLM to automatically verify and assign appropriate materials in bulk. For example, all “window” objects across a large scene can be uniformly updated with the correct electromagnetic profile. This LLM-assisted optimization ensures that the entire scene has consistent and physically accurate material definitions, which is critical for realistic ray-tracing simulation.

The final result is a fully annotated LoD3 indoor or outdoor digital twin with high geometric fidelity and detailed material specifications, ready for channel modeling and simulation.

\subsection{Parameter Definition and Configuration}

Once the environment is modeled, the next step is to configure the simulation parameters. We divide these parameters into two categories: \textit{system parameters}, which define the devices and hardware, and \textit{generation parameters}, which define how the simulation is run and how data is collected.

Below are the main simulation parameter configurations:

\begin{itemize}
    \item \textbf{BS:} BSs are deployed in plausible locations within the scene (e.g., on rooftops or ceilings). Each BS is characterized by attributes including a unique identifier, a 3D position in the scene, an orientation (e.g., given by heading and tilt angles), and a transmit power in dBm.
    
    \item \textbf{MT:} MTs are randomly distributed throughout the environment. Each MT is defined by parameters analogous to the BS. Importantly, we can assign each MT a velocity vector to simulate motion, which introduces temporal variation in the channel, allowing the generation of time-varying CIR as users move.
    
    \item \textbf{Antenna:} Both BSs and MTs are equipped with antenna arrays, defined by their size (number of elements in vertical and horizontal dimensions) and element spacing. These parameters are crucial for accurately modeling the spatial characteristics and beamforming capabilities of the MIMO system.
    
    \item \textbf{RIS:} RIS panels can optionally be included in the environment to intelligently control signal reflections. We consider two beamforming modes:
    
    (i) \textit{Single-beam focusing:} For a point-to-point high-throughput backhaul link, we apply a phase-gradient focusing strategy based on path-length differences to steer and focus the beam. To beamform toward a desired direction $(\theta_0, \phi_0)$, the RIS phase profile is configured as 
    \begin{equation}
    \Phi(x,y) = -\,\frac{2\pi}{\lambda}\,\big(x \sin\theta_0 \cos\phi_0 + \;y \sin\theta_0 \sin\phi_0\big)\,,
    \end{equation}
    assuming the RIS lies in the $xy$-plane. This linear phase gradient across the RIS surface steers a beam in the direction $(\theta_0, \phi_0)$.
    
    (ii) \textit{Multi-beam optimization:} For more complex multi-beam scenarios (e.g., serving multiple users simultaneously), we employ iterative optimization methods (e.g., stochastic gradient descent \cite{zhou2021stochastic}) to design phase codebooks that approximate the desired multi-beam radiation pattern.
    
    \item \textbf{System Bandwidth:} The system bandwidth determines the temporal resolution of the CIR and the frequency sample spacing of the CFR. A wider bandwidth yields higher time resolution (shorter distinguishable multipath delays).
\end{itemize}

\subsection{GPU-Accelerated Ray-Tracing Simulation}

Built upon NVIDIA’s Sionna wireless simulation library, we conduct a GPU-accelerated ray-tracing simulation to generate the channel data. This ray tracer operates on the detailed 3D scenes and uses the configured parameters to propagate signals and capture multipath effects. The process can be broken down into several stages:
\begin{itemize}
    \item \textbf{Initialization:} We launch a multitude of rays from each transmitter. To do this efficiently, we use Fibonacci sphere sampling to distribute rays nearly uniformly across 3D space. If $M$ rays are launched from a transmitter, their directions $(\theta_k, \phi_k)$ can be determined by:
    \begin{equation}
    \theta_k = \arccos\!\Big(1 - \frac{2(k+0.5)}{M}\Big), \qquad 
    \phi_k = \frac{2\pi k}{\Phi_G}\,,
    \end{equation}
    for $k = 0,1,\dots,M-1$, where $\Phi_G = \frac{1+\sqrt{5}}{2}$ is the golden ratio. This parameterization yields an approximately uniform sampling of rays over the sphere. We can also bias the distribution to concentrate rays in critical regions (e.g., denser sampling near ground level where users are), which further improves simulation efficiency.
    
    \item \textbf{Acceleration:} The ray tracing engine is accelerated using NVIDIA’s OptiX framework, enabling highly parallel computation on GPU hardware. Tens of thousands of rays are traced in parallel, each checking for intersections with the environment geometry at every step. This acceleration is crucial for generating a large dataset in a reasonable time, as it allows us to capture rich multipath information without impractical delays.
    
    \item \textbf{Optimization (Early Termination):} To further improve efficiency, we implement two criteria to terminate rays early. Firstly, we set a maximum interaction depth $N_{\max}$ for each ray, counting reflections or refractions. A ray is stopped if it exceeds this limit. Secondly, we impose a power threshold: at each interaction, the ray’s power is attenuated according to material properties and distance traveled. If the remaining power falls below a predefined threshold $P_{\min}$, the ray is terminated, as further interactions would contribute negligible energy. Formally, a ray is terminated if either the number of interactions $N_{\text{int}}$ reaches $N_{\max}$ or its power $P$ drops below $P_{\min}$:
    \begin{equation}
    N_{\text{int}} \ge N_{\max} \quad \text{or} \quad P \le P_{\min}\,. 
    \end{equation}
    This pruning of low-power rays significantly improves simulation speed without affecting accuracy.
    
    \item \textbf{Physics Modeling:} Our ray tracer accounts for all major electromagnetic wave phenomena in a deterministic way. Specifically, we model reflections and transmissions using geometric optics (GO) principles and the Fresnel equations based on material properties. For example, we compute the Fresnel reflection coefficient for each interaction. For perpendicular polarization at an interface between two media, it is given by:
    \begin{equation}
    \Gamma_{\perp}(\theta_i) = \frac{n_1 \cos \theta_i - n_2 \cos \theta_t}{\,n_1 \cos \theta_i + n_2 \cos \theta_t\,}\,,
    \end{equation}
    where $\theta_i$ is the incident angle, $\theta_t$ is the transmitted angle determined by Snell’s law $n_1 \sin\theta_i = n_2 \sin\theta_t$, and $n_1, n_2$ are the refractive indices of the two media. At each interaction, a ray’s complex amplitude $\alpha$ is multiplied by the appropriate reflection ($\Gamma$) or transmission ($T$) coefficient, and by the free-space propagation factor for that path segment. In particular, after a reflection with incident angle $\theta_i$ and propagation distance $d$, the ray’s complex amplitude is updated as:
    \begin{equation}
    \alpha_{\text{new}} = \alpha_{\text{old}}\, \Gamma(\theta_i)\, \frac{e^{-j k d}}{d}\,,
    \end{equation}
    where $k = 2\pi/\lambda$. We also apply the uniform theory of diffraction (UTD) to model how signals bend around sharp edges (e.g., building corners), using standard diffraction coefficients to capture those NLOS paths. To implement these effects efficiently, at each step we perform geometric intersection tests and select the nearest intersection (greedy strategy) for the next interaction. The corresponding physical model of reflection, transmission, or diffraction is then applied at that point, and the precise coordinates and phase of the interaction are logged. All physical parameters for each ray (e.g., phase, amplitude, polarization) are updated as the ray travels and recorded for further processing. 
\end{itemize}

\subsection{Channel Data Extraction and Post-Processing}

After running the ray-tracing simulations, we process and organize the output data into the final dataset format. The DeepTelecom dataset is structured to include both the detailed propagation path data and the synthesized channel responses, as well as auxiliary outputs for visualization. Each scenario (environment) in the dataset yields the following components:

\textbf{Scene Snapshots:} Each scenario is equipped with a freely movable virtual camera, allowing the scene to be captured from various angles and at different time steps. The simulation framework can output rendered images or videos of the environment, which are saved for visualization and analysis.

\textbf{Propagation Paths:} The core output of each scenario is the set of multi-hop propagation paths between every transmitter and receiver. By configuring parameters such as the number of launched rays and the maximum number of reflections/diffractions, the framework records detailed geometric path information. This data is stored as sequences of coordinate points (plus interaction labels), showing the trajectory of signals through the environment. Such path data provides a basis for analyzing propagation mechanisms or optimizing the network (e.g., by placing new relays or RIS).

\textbf{Coverage Heatmaps:} This component generates 2D/3D signal coverage maps for a specified plane in the environment. By configuring the map’s center location, size, orientation (e.g., horizontal plane at a given height), and resolution, the tool calculates and visualizes the received power or path loss at each grid point on that plane. This helps in understanding the impact of obstacles and distance on signal strength, aiding tasks like network planning and deployment optimization.

\textbf{Channel Responses:} Using the physical path data, the dataset provides standard channel information — most importantly the CIR and CFR:
\begin{itemize}
    \item \textbf{CIR:} For a MIMO system with $N_t$ transmit antennas and $N_r$ receive antennas, the time-domain CIR $\mathbf{H}(t)$ is the superposition of contributions from all $L$ propagation paths. Each path $l$ is characterized by its complex gain $\alpha_l$, propagation delay $\tau_l$, AoD $\Omega_{t,l}$, and AoA $\Omega_{r,l}$. The CIR can be written as:
    \begin{equation}
       \mathbf{H}(t) = \sum_{l=1}^{L} \alpha_l\; \mathbf{a}_r(\Omega_{r,l})\; \mathbf{a}_t^H(\Omega_{t,l})\; \delta(t - \tau_l) \,, 
    \end{equation}
    where $\mathbf{a}_r(\Omega_{r,l})$ and $\mathbf{a}_t(\Omega_{t,l})$ are the receive and transmit array response vectors for the path’s AoA and AoD, respectively, and $\delta(\cdot)$ is the Dirac delta.
    
    \item \textbf{CFR:} The CFR is the Fourier transform of the CIR, which is is given by:
    \begin{equation}
        \mathbf{H}(f) = \sum_{l=1}^{L} \alpha_l\; \mathbf{a}_r(\Omega_{r,l})\; \mathbf{a}_t^H(\Omega_{t,l})\; e^{-j 2\pi f \tau_l} \,. 
    \end{equation}
\end{itemize}

\begin{figure}[h]
    \centering
    \includegraphics[width=0.8\linewidth]{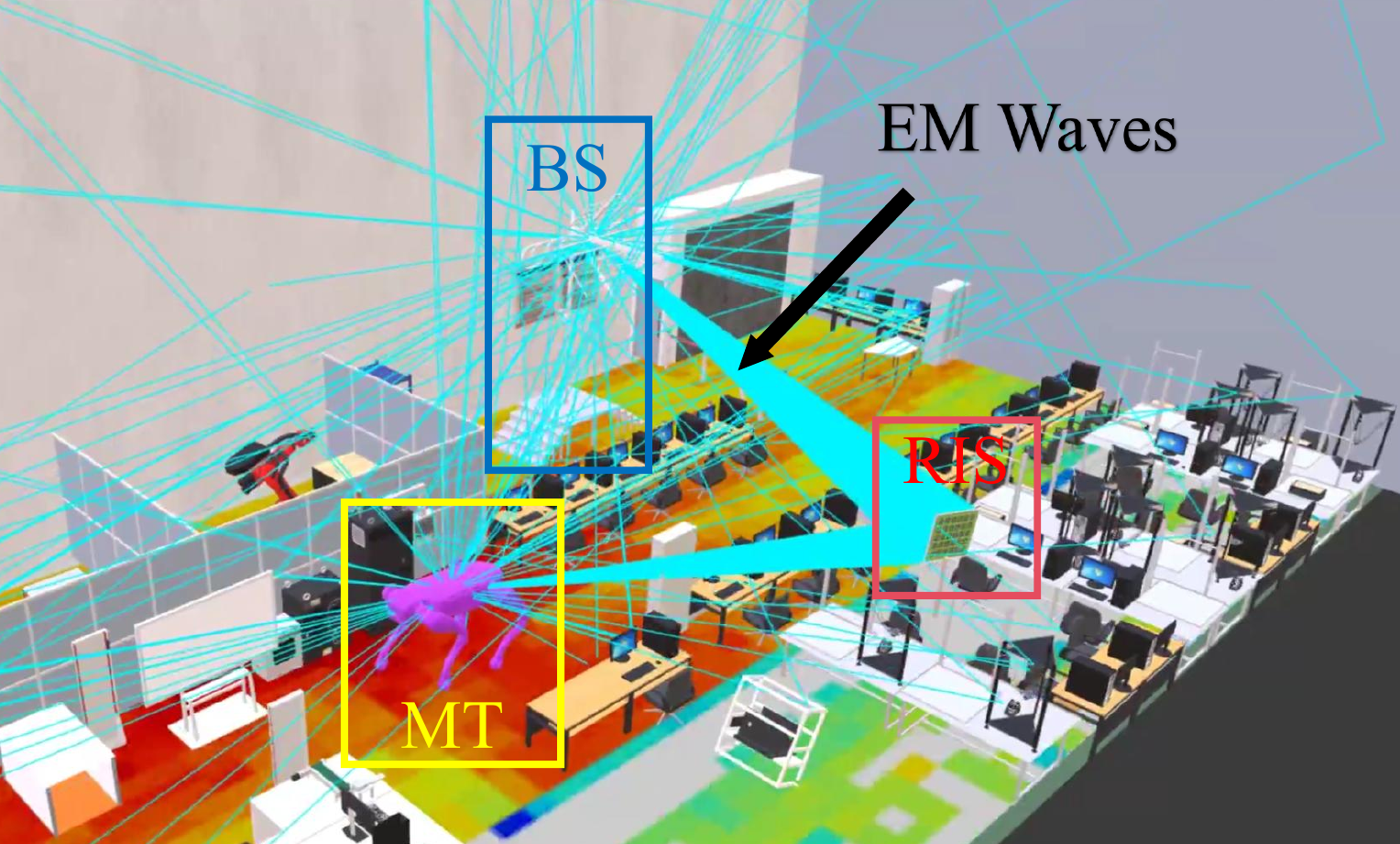}
    \caption{Indoor scene visualization example.}
    \label{fig:visual1}
    \vspace{-2mm}
\end{figure}

\begin{figure}[h]
    \centering
    \includegraphics[width=0.78\linewidth]{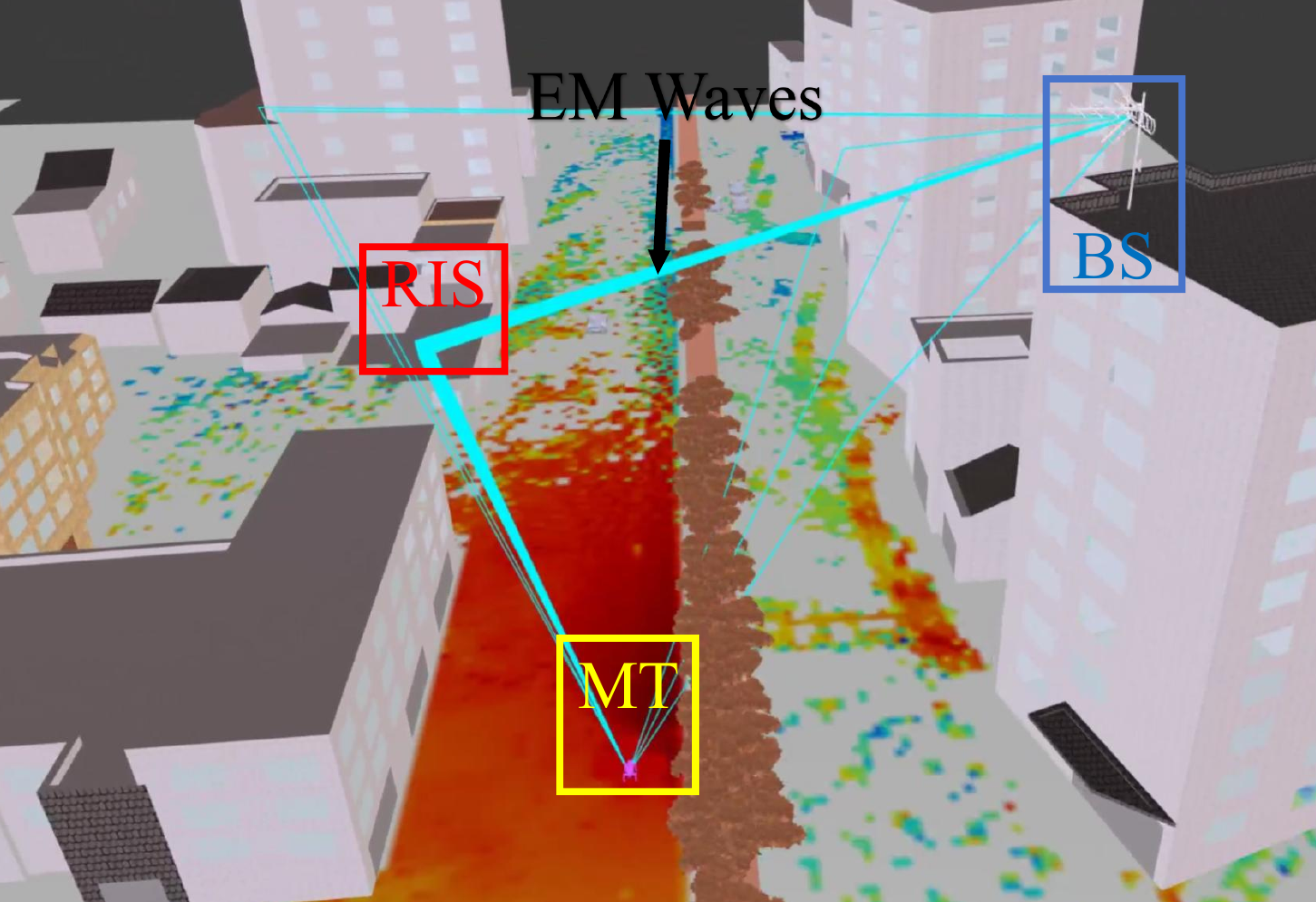}
    \caption{Outdoor scene visualization example.}
    \label{fig:visual2}
    \vspace{-4mm}
\end{figure}

Through this rigorous approach, the DeepTelecom dataset explicitly links the environmental geometry and materials with communication signals, establishing a physically accurate basis for analyzing real-world impacts on MIMO channel behavior in time and frequency domains.
Visualization examples of an indoor and outdoor scene are shown in Figs.~\ref{fig:visual1} and~\ref{fig:visual2}.

\section{Experimental Analysis}
We applied the framework to build the DeepTelecom dataset and assessed its scale and diversity. The results demonstrate the pipeline’s flexibility and the richness of its outputs:
\begin{enumerate}
    \item \textbf{Generality of Scene Generation:} Given GPS coordinates, the system reconstructs outdoor twins with terrain and buildings; given indoor point clouds (e.g., LiDAR or RGB-D), it produces LoD3 interior models. We verified this by generating urban microcell and multi-room office scenarios from their respective geospatial or point-cloud inputs, then deriving channel data for each—showcasing straightforward extensibility to new locales and layouts.
    
    \item \textbf{Comprehensive Scenario Package:} Each scenario is self-contained and includes the 3D model (with materials) and, when needed, an XML scene description, the full simulation configuration (BS/MT placement, antenna/RIS settings), and all outputs: scene snapshots, multipath trajectories, power–delay profiles, coverage heatmaps, and computed CIR/CFR. For example, an outdoor city case comprises the city model, configuration files for BS/MTs, and directories of ray-tracing paths, heatmaps, channel matrices, and accompanying visualizations.
\end{enumerate}

Across dozens of distinct configurations such as downtown, suburban, and rural, each scene yields thousands of transmitter-receiver pairs and millions of raw ray paths. The same pipeline allows users to generate new scenes on demand. The resulting multimodal data support a broad range of wireless AI tasks, including localization \cite{10683036, wang2024multiloc} and beamforming \cite{fenghao1, robustnetwork}, among others.

\section{Discussion and Conclusion}
In this paper, 
we introduce DeepTelecom, a 3D digital-twin channel dataset that leverages a LLM-assisted pipeline to construct detailed outdoor and indoor environments with segmentable and material-parameterizable surfaces at LoD3 precision. Utilizing Sionna’s ray-tracing engine, DeepTelecom accurately simulates comprehensive radio-wave propagation effects. With GPU-accelerated processing, the dataset dynamically generates ray-path trajectories and real-time signal-strength heatmaps, compiling them into high-frame-rate videos while concurrently producing synchronized multi-view images, channel tensors, and multi-scale fading traces. By enabling efficient large-scale, high-fidelity, and multimodal channel data streaming, DeepTelecom not only establishes a standardized benchmark for wireless AI research, but also provides a rich, domain-specific training foundation that facilitates the seamless integration of large model intelligence with next-generation communication systems.




\small
\bibliographystyle{IEEEtran}
\bibliography{bib}
\vspace{12pt}

\end{document}